\documentstyle[manuscript,eqsecnum,aps,version2]{revtex}

\def\xpx{x{\partial\over\partial x}}

\begin{document}
\draft
\preprint{HEP-95-19}
\title
\bf Spherically Symmetric Random Walks III. \\
Polymer Adsorption at a Hyperspherical Boundary
\endtitle

\author{Carl M. Bender}
\instit
Department of Physics, Washington University, St. Louis, MO 63130
\endinstit

\author{Stefan Boettcher}
\instit
Department of Physics, Brookhaven National Laboratory, Upton, NY 11973
\endinstit

\author{Peter N. Meisinger}
\instit
Department of Physics, Washington University, St. Louis, MO 63130
\endinstit

\medskip
\centerline{\today}

\abstract
A recently developed model of random walks on a $D$-dimensional hyperspherical
lattice, where $D$ is {\sl not} restricted to integer values, is used to study
polymer growth near a $D$-dimensional attractive hyperspherical
boundary. The model determines the fraction $P(\kappa)$ of the polymer
adsorbed on this boundary as a function of the attractive potential $\kappa$
for all values of $D$. The adsorption fraction $P(\kappa)$ exhibits a
second-order phase transition with a nontrivial scaling coefficient for
$0<D<4$, $D\neq 2$, and exhibits a first-order phase transition for
$D>4$. At $D=4$ there is a tricritical point with logarithmic scaling. This
model reproduces earlier results for $D=1$ and $D=2$, where $P(\kappa)$
scales linearly and exponentially, respectively. A crossover transition that
depends on the radius of the adsorbing boundary is found.

\endabstract

\section{INTRODUCTION}
\label{s1}
In previous papers \cite{BeBoM,BeCoMe,BeBoMe} we analyzed a class of models of
$D$-dimensional spherically symmetric random walks, where $D$ is {\sl not}
restricted to integer values. In Ref.~\cite{BeBoM} we introduced the notion of
spherically symmetric random walks and in Ref.~\cite{BeCoMe} we studied a
simplified model of spherically symmetric random walks that is analytically
tractable for all values of $D$. In Ref.~\cite{BeBoMe} we considered random
walks that allow for the creation and annihilation of random walkers and
demonstrated that these extended models exhibit critical behavior as a function
of the birth rate of walkers. This critical behavior exhibits an interesting
dependence on the dimension $D$. In this paper we apply these ideas to the study
of polymer growth near a $D$-dimensional hyperspherical adsorbing boundary.

Polymers have inspired many experimental and theoretical investigations
\cite{AIP,DesCJ}. Because polymers are complex objects constructed from
simple building blocks, they serve as a laboratory for the development of
scaling methods \cite{DeG}, renormalization group theory \cite{Freed}, and Monte
Carlo simulation \cite{Binder}. Formulating simplified statistical models of
polymer growth is useful for understanding aspects of critical phenomena
exhibited by actual polymers. Indeed, any solvable statistical models that
exhibit nontrivial critical behavior are worthy of study \cite{Bax}.

The simplest polymer system is an unbranched chain of monomers. Such systems are
easy to model by means of self-avoiding random walks \cite{FloMo,Sokal}. In this
paper, we examine such a polymer growing in the neighborhood of an attractive
$D$-dimensional hyperspherical boundary. Special cases of this polymer system
have already been investigated for planar ($D=1$) \cite{Priv} and cylindrical
($D=2$) \cite{BoMo,Bo} boundaries.

The spherically symmetric random-walk model introduced in Ref.~\cite{BeBoM} was
used in Ref.~\cite{BoMo} to determine the critical properties of a polymer
growing near an attractive $D$-dimensional hyperspherical boundary. However,
that model was mathematically intractable except for $D=1$ and $D=2$. Here, we
use the {\sl new\/} model of hyperspherical random walks introduced in
Ref.~\cite{BeCoMe} to solve the $D$-dimensional polymer growth model for
arbitrary $D>0$. Specifically, we consider an ensemble of polymers near an
attractive $D$-dimensional hyperspherical boundary of radius $m$, where $m
\geq 0$ is measured in discrete monomer units. We derive the adsorption fraction
$P(\kappa)$ as a function of the attractive potential $\kappa$ in the limit
where the average length of a polymer reaches infinity. The parameter $\kappa$
is closely related to the birth rate $a$ used in Ref.~\cite{BeBoMe}.

We find that if the attractive potential $\kappa$ drops below a critical value
$\kappa_{\rm c}$, which in general depends on $D$, the adsorption fraction
vanishes. As $\kappa-\kappa_{\rm c}\to 0^+$ for fixed radius $m$, the asymptotic
behavior of $P(\kappa)$ is given by
\begin{eqnarray}
P(\kappa)\sim \cases{
{\rm C}_1(D,m) \left(\kappa-\kappa_{\rm c}\right)^{D\over 2-D} &$(0<D<2),$\cr
\noalign{\medskip}
{\rm C}_2(2,m)(\kappa-\kappa_{\rm c})^{-2}{\rm exp}\left [\displaystyle{-{8\over
9(m+1)(\kappa-\kappa_{\rm c})}}\right]& $(D=2),$\cr
\noalign{\medskip}
{\rm C}_3(D,m) \left(\kappa-\kappa_{\rm c}\right)^{4-D\over D-2} &$(2<D<4),$\cr
\noalign{\medskip}
{\rm C}_4(4,m){\displaystyle {1\over\ln (\kappa-\kappa_{\rm c})}} &$(D=4),$\cr
\noalign{\medskip}
{\rm C}_5(D,m) &$(D>4),$\cr}
\label{e1.1}
\end{eqnarray}
where ${\rm C}_i(D,m)$ are constants that depend on the dimension $D$ and the
radius $m$ of the adsorbing boundary.

Equation (\ref{e1.1}) ceases to be valid as $m\to\infty$. When
\begin{eqnarray}
\kappa-\kappa_{\rm c} \sim {B(D)\over m},
\label{e1.2}
\end{eqnarray}
where $B(D)$ is a constant of order 1, we observe a {\sl crossover} transition
to linear scaling behavior in $P(\kappa)$ as $\kappa-\kappa_{\rm c}\to 0^+$.

In Sec.~\ref{s2} we discuss the theory of polymer growth near an attractive
$D$-dimensional spherically symmetric boundary. In Sec.~\ref{s3} we solve
the eigenvalue problem that results from a transfer matrix description of
this growth process. Finally, in Sec.~\ref{s4} we determine for all $D>0$
the adsorption fraction $P(\kappa)$ near the critical point $\kappa_{\rm c}$.

\section{DIRECTED WALK MODEL FOR POLYMER ADSORPTION}
\label{s2}
We model polymer growth as a nonintersecting (directed), random walk in
$D+1$-dimensional space. This random walk takes place on the union of a 
one-dimensional semi-infinite lattice and a $D$-dimensional lattice consisting
of a set of concentric hyperspherical surfaces labeled $S_n$. The hyperspherical
surfaces are equally spaced in units of one monomer length. The innermost
surface $S_m$, $m\geq 0$, is the attractive boundary, which has a radius of $m$
in monomer units. The next surface $S_{m+1}$ has a radius of $m+1$, and so on.
The extra axial dimension is introduced to ensure that the random walk is
nonintersecting; thus, we are actually studying a cylindrically-symmetric
random walk in $D+1$ dimensions.

At each step the random walker has a probability of moving one monomer unit
radially outward, moving one monomer unit radially inward, or staying on the
same radial surface. (When the walker is on the boundary surface $S_m$ the 
walker's probability of moving inward is zero.) Regardless of whether the walker
moves radially or remains on the same radial surface, we then require the walker
to move one additional monomer unit in the axial direction in the $D+1$st
dimension. This deterministic axial motion guarantees that the random walk will
never cross itself. (A similar requirement is imposed in restricted
solid-on-solid models.) Hence, at each step the polymer grows by adding either
one or two monomers, but always advances exactly one unit in the axial
direction. The growth of such a polymer is illustrated in Fig.~\ref{f1} for
the case $D=2$.

The dynamics of the polymer growth is regulated by a balance between energy and
entropy. There is one energy associated with the addition of a new monomer and
another associated with adsorption on the attractive boundary. Each addition of
a monomer is characterized by a factor of $z$ and each addition of a monomer on
the attractive boundary $S_m$ is associated with an additional factor of
$\kappa$. The factor $\kappa$ is shown in Fig.~\ref{f1} but the factor of $z$ is
not indicated because there is one such factor for each line segment (monomer).
As the dimension $D$ increases there is a corresponding increase in the
available volume for the polymer to occupy as it grows away from the adsorbing
boundary. For any given $D$ this configurational entropy balances the binding
potential on the attractive boundary. Thus, one might anticipate that the
critical properties of this system will vary in an interesting way as a function
of the curvature of the boundary.

While this random walk model is only a crude description of an actual polymer
growing in a continuum, one might hope that the critical properties of the
polymer system in the infinite chain limit are universal and well approximated
by such a model.

We consider next the probabilities that define the radial motion of the random
walk. We have introduced a hyperspherical lattice because spherical symmetry
reduces a $D$-dimensional problem to a one-dimensional problem. The probability
distribution of a spherically symmetric random walk is described completely by a
one-dimensional recursion relation.\cite{BeBoM} The coefficients in this
recursion relation are dependent on the location of the walker and express the
radial bias (or entropy) of the spherical geometry; that is, a random walker
tends to move outward rather than inward because more volume is available in the
outward direction when $D>1$. In Ref.~\cite{BeBoM} inward and outward walk
probabilities were proposed that express this radial bias. Unfortunately, for
arbitrary $D$, these probabilities are so complicated that an analytical
solution to the recursion relation is impossible except for a few special values
of $D$.

In a recent paper \cite{BeCoMe} it was shown that the recursion relation
{\sl can\/} be solved analytically for all $D>0$ by replacing the outward and
inward walk probabilities for region $S_n$ with a uniform approximation for all
$n$. In Ref.~\cite{BeCoMe} it was shown that this simplified random walk
exhibits the usual scaling properties of a random walk model. For example, walks
on this lattice have a Hausdorff dimension $D_H=2$. In comparison with random
walks on other lattices, such as a hypercubic lattice, the random walk model
studied in this paper is remarkable because it is analytically tractable.
Numerical and analytical studies in Ref.~\cite{BeBoMe} suggest that, despite the
simplicity of the model, the nontrivial phenomena obtained in this paper are
indeed universal. 

We represent the probabilities that define the random walk considered in this
paper by $P_{\rm stay}(n)$, the probability that a walker stays on the surface
$S_n$ and just moves in the axial direction, $P_{\rm out}(n)$, the probability
that the walker moves outward from the surface $S_n$ to the surface $S_{n+1}$
(and then moves in the axial direction on the surface $S_{n+1}$), and
$P_{\rm in}(n)$, the probability that the walker moves inward from the surface
$S_n$ to the surface $S_{n-1}$ (and then moves in the axial direction on the
surface $S_{n-1}$). Generalizing the probabilities used in Ref.~\cite{BeCoMe} to
include the possibility of staying on the surface $S_n$, we express the relative
probabilities as
\begin{eqnarray}
P_{\rm stay}(n)\equiv 1,\quad P_{\rm in}(n)={2n\over {2n+D-1}}, \quad P_{\rm
out}(n)={2(n+D-1)\over {2n+D-1}}\quad (n>m).
\label{e2.1}
\end{eqnarray}
Note that the walker is more likely to move outward as $D$ increases. However,
as $m$ increases with $D$ held fixed, the outward and inward probabilities
become equal; this happens because at large radius our nested spheres appear
locally (on the scale of a monomer length) to be equally spaced parallel planes.

On the boundary $S_m$ we enforce the condition that the walker is prohibited
from moving inward by requiring that 
\begin{eqnarray}
P_{\rm stay}(m)=1,\quad P_{\rm in}(m)=0, \quad P_{\rm out}(m)=1. 
\label{e2.2}
\end{eqnarray}
The probabilities for the special one-dimensional case considered in
Ref.~\cite{Priv} are obtained if we set $D=1$ in Eqs.~(\ref{e2.1}).

A single random walk can represent only one of the many configurations that a
growing polymer can attain. To obtain the critical properties of polymer growth
at an attractive boundary, we must investigate the average behavior of an
ensemble of walkers. Thus, we derive a partition function for random walks and
use it to generate ensemble averages that describe, for example, the fraction of
a growing polymer that is adsorbed on the boundary. From the above probabilities
and the parameters $z$ and $\kappa$ associated with adding monomers, we
construct a transfer matrix $T_{j,i}$ that expresses the probability of the
walker moving from the $i$th to the $j$th surface at each step: 
\begin{eqnarray}
T_{j,i}=z^{\vert j-i\vert} \kappa^{\delta_{m,j}}\left [P_{\rm stay}(i)\delta_{j,
i}+P_{\rm out}(i)\delta_{j-1,i}+P_{\rm in}(i)
\delta_{j+1,i}\right ]. 
\label{e2.3}
\end{eqnarray}

A particular polymer configuration generated by a random walk consisting of $L$
steps is characterized by a set of $L$ integers $\{h_i\}_{i=1}^{L}$ that specify
the surface $S_{h_i}$ reached on the $i$th step in the axial direction. The
total statistical weight of such a polymer is expressed as a product of $L$
elements of the transfer matrix:
\begin{eqnarray}
z^L \delta_{m,h_0} T_{h_1,h_0} T_{h_2,h_1}\ldots T_{h_L,h_{L-1}},
\label{e2.4}
\nonumber
\end{eqnarray}
where the Kronecker delta ensures that the polymer is initially grafted to the
boundary. The partition function $Z_L$ for all polymers having axial length $L$
is then 
\begin{eqnarray}
Z_L=z^L {\vec b}^{(t)} T^L {\vec e},
\label{e2.5}
\nonumber
\end{eqnarray}
where ${\vec b}^{(t)}$ and ${\vec e}$ are vectors accounting for beginning
and end effects. Hence, $Z=\sum_{L=1}^{\infty} Z_L$, the total partition
function for configurations of all axial lengths, is given by
\begin{eqnarray}
Z(z,\kappa)={\vec b}^{(t)} zT (1-zT)^{-1} {\vec e}.
\label{e2.6}
\end{eqnarray}

Let $\lambda_{\rm max}(\kappa,z)$ be the largest eigenvalue of the transfer
matrix $T$ and define $z_{\infty}(\kappa)$ by
\begin{eqnarray}
1=z_{\infty}(\kappa)\lambda_{\rm max}[\kappa,z_{\infty}(\kappa)].
\label{e2.7}
\end{eqnarray}
Letting $\Delta z=z_\infty(\kappa)-z$, note that the partition function $Z$ in
Eq.~(\ref{e2.6}) diverges as $\Delta z\to 0^+$.

We can express the average length of a polymer in terms of the partition
function $Z$:
\begin{eqnarray}
\langle N(z,\kappa)\rangle=z {\partial\over\partial z} \ln{Z(z,\kappa)}.
\label{e2.8}
\nonumber
\end{eqnarray}
Similarly, the average number of monomers adsorbed on the boundary is given by
\begin{eqnarray}
\langle N_{S_m}(z,\kappa)\rangle= \kappa {\partial\over\partial\kappa} 
\ln{Z(z,\kappa)}. 
\label{e2.9}
\nonumber
\end{eqnarray}
As $\Delta z\to 0^+$, both $\langle N\rangle$ and $\langle N_{S_m}\rangle$ 
diverge; that is, the average length of a polymer chain diverges. In this paper
we study the fraction of adsorbed monomers $P(\kappa)$ as a function of the
binding potential $\kappa$ for an ensemble of polymers of all possible lengths.
The adsorption fraction is given by
\begin{eqnarray}
P(\kappa)=\lim_{\Delta z\to 0^+}{\langle
N_{S_m}(z,\kappa)\rangle\over\langle N(z,\kappa)\rangle }=-{\kappa\over
z_{\infty}(\kappa)}{d z_{\infty}(\kappa) \over d\kappa}.
\label{e2.10}
\end{eqnarray}
Note that the adsorption fraction $P(\kappa)$ is defined only on the line
$z_{\infty}(\kappa)$ in the $(\kappa,z)$-plane. 

\section{EIGENVALUES OF THE TRANSFER MATRIX FOR ARBITRARY $D$} 
\label{s3}
In this section we use generating-function techniques to derive a differential
equation whose solution yields the spectrum $\lambda$ of the transfer matrix $T$
defined in Eq.~(\ref{e2.3}). We begin by inserting the probabilities in
Eqs.~(\ref{e2.1}-\ref{e2.2}) into Eq.~(\ref{e2.3}) and obtain the
difference-equation eigenvalue problem \cite{Tr}
\begin{eqnarray}
\lambda g_n=\sum_{i=m+1}^{\infty} T_{n,i} g_i =\cases{ g_n+2z {n+D-2\over{2n+D-3
}}g_{n-1}+2z {n+1\over {2n+D+1}} g_{n+1}& ($n\geq m+2$),\cr
\noalign{\medskip}
g_{m+1}+z g_m+2z {m+2\over {2m+D+3}} g_{m+2}& ($n=m+1$),\cr
\noalign{\medskip}
\kappa g_m+2\kappa z {m+1\over {2m+D+1}} g_{m+1}& ($n=m$).\cr} 
\label{e3.1}
\end{eqnarray}

This problem has a continuous spectrum for all values of $\kappa$, but the
spectrum contains bound states only for a certain range of $\kappa$. The
continuous spectrum, and thus the value of its upper limit, does not vary as a
function of $\kappa$. If $\kappa$ is in a range such that the upper limit of the
continuous spectrum is the largest eigenvalue of the transfer matrix, the
adsorption fraction as defined in Eq.~(\ref{e2.10}) vanishes because, by the
chain rule, it is proportional to the derivative of $\lambda_{\rm max}$ as a
function of $\kappa$. On the other hand, if the value of $\kappa$ is such that a
bound state exists, the bound-state eigenvalue usually {\sl does} vary as a
function of $\kappa$, and its value is larger than the upper limit of the
continuous spectrum, which leads to a nonvanishing adsorption fraction. Thus,
the emergence of bound states is the criterion for the appearance of an adsorbed
phase for the polymer.

A bound state must satisfy a condition ensuring that the likelihood of finding
the walker in remote regions $n\to\infty$ is diminishing sufficiently fast:
\begin{eqnarray}
g_n\to 0\qquad (n\to\infty).
\label{e3.1.1}
\end{eqnarray}
To find such a condition it is convenient to define
\begin{eqnarray}
g_n=\cases{\left(2n+D-1\right) h_n& ($n>m$),\cr 
\noalign{\medskip}
2(m+D-1) h_m& ($n=m$).\cr}
\label{e3.2}
\nonumber
\end{eqnarray}
Then, Eqs.~(\ref{e3.1}) reduce to
\begin{eqnarray}
0=(1-\lambda)\left(2n+D-1\right) h_n+2z \left(n+D-2\right)
h_{n-1}+2z (n+1) h_{n+1}\quad (n>m),
\label{e3.3}
\end{eqnarray}
supplemented by the boundary condition
\begin{eqnarray}
0=(\kappa-\lambda)(m+D-1) h_m + \kappa z(m+1) h_{m+1}.
\label{e3.4}
\end{eqnarray}

To simplify the analysis of this problem, we define the generating functions 
\begin{eqnarray}
G(x)=\sum_{n=m}^{\infty} g_n x^n
\label{e3.5}
\nonumber
\end{eqnarray}
and
\begin{eqnarray}
H(x)=\sum_{n=m}^{\infty} h_n x^n.
\label{e3.6}
\nonumber
\end{eqnarray}
Using the identity
\begin{eqnarray}
\sum_n n x^n h_n = \xpx \sum_n x^n h_n,
\label{e3.7}
\end{eqnarray}
$G(x)$ can be formally obtained from $H(x)$:
\begin{eqnarray}
G(x)=2\xpx H(x)+(D-1)\left[H(x)+ x^m h_m\right].
\label{e3.8}
\end{eqnarray}

A differential form of the eigenvalue problem may now be obtained by multiplying
Eq.~(\ref{e3.3}) by $x^n$ and summing from $n=m+1$ to $n=\infty$. After shifting
indices and applying the identity in Eq.~(\ref{e3.7}), we obtain
\begin{eqnarray}
&&\left[(1-\lambda)\left(D-1+2\xpx\right)+2zx\left(D-1+\xpx\right)+
2z{\partial\over\partial x}\right] H(x)\cr
\noalign{\medskip}
&=&\left[(1-\lambda)(2m+D-1)+m{2z\over x}\right]h_m x^m+2z(m+1) h_{m+1} x^m.
\label{e3.9}
\nonumber
\end{eqnarray} 
We eliminate $h_{m+1}$ by applying the boundary condition in Eq.~(\ref{e3.4})
and divide both sides by $2z$ to obtain
\begin{eqnarray}
Q^2(x) H'(x) + (D-1) Q(x) Q'(x) H(x)=\left(1+Ax\right) m h_m x^{m-1},
\label{e3.11}
\end{eqnarray}
where we have defined
\begin{eqnarray}
A={m+D-1\over m} \left[ {1\over\kappa}\left({2\over\epsilon} +
{1\over z}\right) - {1\over z}\right] - {2m+D-1\over m\epsilon}
\label{e3.12}
\end{eqnarray}
and
\begin{equation}
\epsilon={2z\over \lambda-1},\quad \gamma={1\over\epsilon} 
\left( 1-\sqrt{1-\epsilon^2}\right),\quad Q(x)=\sqrt{(x-\gamma)(x-1/\gamma)}.
\label{e3.10}
\end{equation}

It is easy to solve Eq.~(\ref{e3.11}) because it is a linear first-order
differential equation. We multiply by the integrating factor $Q(x)^{D-3}$ to get
\begin{eqnarray}
\left[Q^{D-1}(x) H(x)\right]'=(1+Ax) Q^{D-3}(x) m h_m x^m.
\label{e3.13}
\nonumber
\end{eqnarray}
Requiring that
$$\lim_{x\to 0} x^{-m}H(x)=h_m$$
gives
\begin{eqnarray}
H(x)=m h_m Q^{1-D}(x)\int_0^x dt~t^{m-1} Q^{D-3}(t)(1+A t).
\label{e3.14}
\nonumber
\end{eqnarray}
Thus, from Eq.~(\ref{e3.8}) the generating function $G(x)$ is given by
\begin{eqnarray}
G(x)&=&(D-1)m h_m \Bigg\{ x^m\left[1+{2(1+Ax)\over (D-1)Q^2(x)}\right]
\nonumber\\
&&\quad +{(1-x^2)\over Q^{D+1}(x)}\int_0^x dt\,t^{m-1}Q^{D-3}(t)(1+At)\Bigg\}.
\label{e3.15}
\end{eqnarray}

The behavior of $g_n$ as $n\to\infty$ is determined by the singularities of
$G(x)$. It is evident from the definition of $Q(x)$ in Eqs.~(\ref{e3.10}) that
$G(x)$ in general has singularities at $x=\gamma$ and $x=1/\gamma$. If $\gamma$
is complex, then $\gamma=1/\gamma^*$, and both singularities are located on the
unit circle. Thus, condition (\ref{e3.1.1}) cannot be satisfied and there is no
bound state. The largest value of the transfer matrix is given by the upper
limit of the continuous spectrum, $\lambda_{max}(\kappa)=2$. Hence, the
adsorption fraction vanishes and the polymer is in the desorbed phase.

To obtain a nonzero adsorption fraction $P(\kappa)$ we must find bound states in
the spectrum $\lambda$. Bound states (discrete values of $\lambda$) appear for
values of $\kappa$ and $z$ such that $\gamma$ is real and $\gamma<1$. We must
eliminate growing solutions of the form $g_n\propto\gamma^{-n}$. This is
accomplished by imposing the finiteness condition  
\begin{eqnarray}
\lim_{x\to\gamma} |G(x)|<\infty.
\label{e3.16}
\nonumber
\end{eqnarray}

A local analysis of $G(x)$ for $x\to\gamma^-$ reveals that $G(x)$ is finite at 
$x=\gamma$ if the following eigenvalue condition is satisfied:
\begin{eqnarray}
0=\int_0^1 dt~t^{m-1}(1+A\gamma t)\left[(1-t)(1-\gamma^2 t)\right]^{D-3\over 2}.
\label{e3.17}
\end{eqnarray}
This integral is divergent for $D\leq 1$ (or for $D\leq 2$ when $\gamma=1$).
Therefore, to study this integral for all values of $D$ we observe that when
it converges it defines a hypergeometric function ${}_2F_1(a,b;c;z)$ \cite{A+S}.
We then rely on the analytic continuation provided by the hypergeometric
function to rewrite Eq.~(\ref{e3.17}) as
\begin{eqnarray}
0={2m+D-1\over 2m}{}_2F_1\left({3-D\over 2},m;m+{D-1\over 2};\gamma^2\right)  
+ \gamma A {}_2F_1\left({3-D\over 2},m+1;m+{D+1\over 2};\gamma^2\right).
\label{e3.18}
\nonumber
\end{eqnarray}
Next we use the quadratic transformation formula for hypergeometric functions
15.3.26 in Ref.~\cite{A+S} to obtain 
\begin{eqnarray}
0&=&{2m+D-1\over m}{}_2F_1\left({m\over 2},{m+1\over 2};m+{D-1\over 2};\epsilon
^2\right)\cr
\noalign{\bigskip}
&&\qquad\qquad\qquad +\epsilon A~{}_2F_1\left({m+1\over 2},{m\over 2}+1;m+{D+1
\over 2}; \epsilon^2\right).
\label{e3.19}
\end{eqnarray}

Using $z=z_{\infty}(\kappa)$, $A$ and $\epsilon$ as defined in
Eqs.~(\ref{e3.9}) and (\ref{e3.10}), and Eq.~(\ref{e2.7}), the eigenvalue
condition in (\ref{e3.19}) yields an implicit relation for $z_{\infty}(\kappa)$,
and thus we obtain the adsorption fraction $P(\kappa)$ as defined in
Eq.~(\ref{e2.10}).

\section{CRITICAL POINT ANALYSIS}
\label{s4}
The numerical value of $P(\kappa)$ for any $\kappa$ can be obtained from the
implicit equation for $z_{\infty}(\kappa)$. However, using asymptotic analysis
we can determine the behavior near the critical point explicitly. We showed that
the critical transition is associated mathematically with the onset of bound
states. Thus, the critical point is located at $\epsilon=1$. Using
\cite{A+S}
\begin{eqnarray}
{}_2F_1(a,b;c;x)&=&{\Gamma(c)\Gamma(c-a-b)\over\Gamma(c-a)\Gamma(c-b)}
{}_2F_1(a,b;a+b-c+1;1-x)\cr
\noalign{\medskip}
&&+~(1-x)^{c-a-b} {\Gamma(c)\Gamma(a+b-c)\over\Gamma(a)\Gamma(b)}{}_2F_1(c-a,c-
b;c-a-b+1;1-x),
\label{e4.1}
\nonumber
\end{eqnarray}
we rewrite the eigenvalue condition (\ref{e3.19}) as
\begin{eqnarray}
&K&{m+1\over m+D-1}\left({1-\epsilon^2\over 12}\right)^{{D\over 2}-1}
\Bigg [ {}_2F_1\left({m+D-1\over 2},{m+D-2\over 2};{D\over 2};1-\epsilon^2
\right)\nonumber\\
&&\qquad\qquad+\epsilon A~{}_2F_1\left({m+D\over 2},{m+D-1\over 2};{D\over 2};1-
\epsilon^2\right)\Bigg ] \nonumber\\
&=&{m+D-2\over m}{}_2F_1\left({m\over 2},{m+1\over 2};{4-D\over 2};1-\epsilon^2
\right)\nonumber\\
&&\qquad\qquad+\epsilon A~{}_2F_1\left({m+1\over 2},{m\over 2}+1;{4-D\over 2};1-
\epsilon^2\right)
\label{e4.2}
\end{eqnarray}
with
\begin{eqnarray}
K=-3^{{D\over 2}-1} {\Gamma\left(1-{D\over 2}\right)\Gamma\left(m+D\right)
\over \Gamma\left({D\over 2}-1\right)\Gamma(m+2)}.
\label{e4.2.1}
\end{eqnarray}
Observe that the special cases of even integer $D>2$ require special attention;
we consider these special cases later. The case $D=2$ has already been studied
in Ref.~\cite{Bo} and will not be discussed here.

We now substitute
\begin{eqnarray}
z&=&z_{\infty}(\kappa_{\rm c})-\Delta z\quad (\Delta z\to 0^+),\cr
\noalign{\medskip}
\kappa&=&\kappa_{\rm c}+\Delta \kappa\quad (\Delta\kappa\to 0^+). 
\label{e4.3}
\nonumber
\end{eqnarray}
We will determine $\kappa_{\rm c}$ later from asymptotic analysis of the
eigenvalue condition. However, using $\epsilon\vert_{z_{\infty}(\kappa_{\rm c})}
=1$ and $\lambda=1/z_{\infty}(\kappa)$ and the definition of $\epsilon$ in
Eqs.~(\ref{e3.10}), is easy to determine that
\begin{eqnarray}
z_{\infty}(\kappa_{\rm c})={1\over 2}.
\label{e4.4}
\nonumber
\end{eqnarray}
Retaining terms to sufficient order for the subsequent analysis,
\begin{eqnarray}
\epsilon&\sim&1-6\Delta z+\ldots,\cr
\noalign{\medskip}
A&\sim& {m+D-1\over m}\left[\left({4\over\kappa_{\rm c}}-3-{m\over m+D-1}\right)
-{4\over\kappa_{\rm c}^2}\Delta\kappa+2\left({8\over\kappa_{\rm c}}-
{8m+5D-5\over m+D-1}\right) \Delta z+\ldots\right],
\label{e4.5}
\nonumber
\end{eqnarray}
we obtain from the eigenvalue condition in (\ref{e4.2})
\begin{eqnarray}
{\cal A}~+~{\cal B}\,\Delta\kappa~+~{\cal C}\,\Delta z\ldots~\sim~\Delta z^{{D
\over 2}-1}\,\left( {\cal X}\,+\,{\cal Y}\,\Delta\kappa+\ldots\right),
\label{e4.6}
\end{eqnarray}
where
\begin{eqnarray}
{\cal A}&=&3-{4\over\kappa_{\rm c}}-{D-2\over m+D-1},\cr
\noalign{\medskip}
{\cal B}&=&4{m+D-1\over (m+1)\kappa_{\rm c}^2},\cr
\noalign{\medskip}
{\cal C}&=& {2\over (m+1)(4-D)} \bigm[2(D-1)(2D+1)+m(28D-13)+6m^2(D+5)+9m^3\cr
\noalign{\medskip}
&&\qquad\qquad\qquad\qquad -{4\over\kappa_{\rm c}}(m+D-1)(3m^2+9m+D+2)\bigm],\cr
\noalign{\bigskip}
{\cal X}&=& \left(3-{4\over\kappa_{\rm c}}\right) K,\cr
\noalign{\medskip}
{\cal Y}&=& {4K\over\kappa_{\rm c}^2},
\label{e4.7}
\nonumber
\end{eqnarray}
where, $K$ is given in Eq.~(\ref{e4.5}).

In the following subsections, we determine the critical point $\kappa_{\rm c}$
and the asymptotic relation between $\Delta z$ and $\Delta\kappa$ by balancing
terms in Eq.~(\ref{e4.6}) order by order in the limit $\Delta z\to 0^+,~\Delta
\kappa\to 0^+$. The asymptotic behavior of the adsorption transition near the
critical point is then obtained from
\begin{eqnarray}
P(\kappa)\sim {\kappa_{\rm c}\over z_{\infty}(\kappa_{\rm c})}{d\Delta
z\over d\Delta\kappa}.
\label{e4.8}
\end{eqnarray}

\subsection{Case $0<D<2$}
\label{ss4.1}
In this case we eliminate a divergent term in Eq.~(\ref{e4.6}) by imposing
the condition ${\cal X}=0$, which gives
\begin{eqnarray}
\kappa_{\rm c}={4\over 3}.
\label{e4.9}
\end{eqnarray}
To balance the terms in next order we demand that
\begin{eqnarray}
{\cal A}\sim {\cal Y} \Delta z^{{D\over 2}-1} \Delta\kappa.
\label{e4.10}
\nonumber
\end{eqnarray}
Thus, we find that
\begin{eqnarray}
\Delta z\sim \left({{\cal Y}\over{\cal A}}\right)^{2\over 2-D}\,
\Delta\kappa^{2\over 2-D}
\label{e4.11}
\nonumber
\end{eqnarray}
and, according to Eq.~(\ref{e4.8}),
\begin{eqnarray}
P(\kappa)\sim {16\over 3(2-D)} \left[{9(m+1) \over 4(2-D)}K\right]^{2\over 2-D}
(\kappa-\kappa_{\rm c})^{D\over 2-D}\quad (\kappa\to\kappa_{\rm c}^+). 
\label{e4.12}
\end{eqnarray}

\subsection{Case $2<D<4$}
\label{ss4.2}
Here, to balance the most dominant terms in Eq.~(\ref{e4.6}) we set
${\cal A}=0$, which yields
\begin{eqnarray}
\kappa_{\rm c}={4\over 3-{D-2\over m+D-1}}.
\label{e4.13}
\end{eqnarray}
Note that $\kappa_{\rm c}$ in Eq.~(\ref{e4.13}) joins continuously onto the
value of $4/3$ in Eq.~(\ref{e4.9}) at $D=2$ for all $m$, rises with $D$ for
$D>2$, and levels off at $\kappa_{\rm c}=2$ as $D\to\infty$. On the other hand,
as $m\to\infty$ for fixed $D>2$ we regain the value of the critical point for
$D=1$. 

To next order in the asymptotic analysis we obtain
\begin{eqnarray}
\Delta z~\sim~\left({{\cal B}\over{\cal X}}\right)^{2\over D-2}
\Delta\kappa^{2\over D-2},
\label{e4.14}
\nonumber
\end{eqnarray}
or
\begin{eqnarray}
P(\kappa) &\sim& {16(m+D-1)\over (D-2)(3m+2D-1)} \left[{(3m+2D-1)^2
\over 4K(D-2)(m+1)} \right]^{2\over D-2}  (\kappa-\kappa_{\rm c})^{4-D\over D-2}
\quad (\kappa \to\kappa_{\rm c}^+).
\label{e4.15}
\end{eqnarray}

\subsection{Case $D>4$}
\label{ss4.3}
As in the previous subsection we must set ${\cal A}=0$ and obtain the same value
for $\kappa_{\rm c}$ as in Eq.~(\ref{e4.13}). Higher-order asymptotic analysis
then gives
\begin{eqnarray}
\Delta z\sim -{{\cal B}\over{\cal C}} \Delta\kappa.
\label{e4.16}
\nonumber
\end{eqnarray}
Thus, to leading order the adsorption fraction is asymptotically a constant:
\begin{eqnarray}
P(\kappa)\sim {(D-4)(3m+2D-1)\over (2m+D)(3m+2D-5)}\quad(\kappa\to
\kappa_{\rm c}^+).
\label{e4.17}
\end{eqnarray}
The discontinuity in the adsorption fraction across the critical point
indicates a first-order phase transition.

The higher-order correction to this jump discontinuity is given by a term of
order $\Delta\kappa^{{D\over 2}-2}$. This correction dominates for $D<6$. When
$D\geq 6$, the dominant correction becomes a term of order $\Delta\kappa$,
scaling independently of $D$. Note that the jump discontinuity disappears as
$m\to\infty$.

\subsection{Special Case $D=4$}
\label{ss4.4}
The case $D=4$ is special because here a line of first and second-order phase
transitions meet. Our analysis must proceed somewhat differently because certain
coefficients in Eq.~(\ref{e4.6}) diverge as $D\to 4$. To investigate this
case we return to Eq.~(\ref{e3.19}), evaluated at $D=4$, and use the relation
(see formula 15.3.11 in Ref.~\cite{A+S})
\begin{eqnarray}
{}_2F_1\left({m\over 2},{m+1\over 2};m+{3\over 2};\epsilon^2\right)\sim
{2^{m+1}\Gamma\left(m+{3\over 2}\right)\over \sqrt{\pi}\Gamma(m+2)}
\left[1+3m(m+1)\Delta z \ln{\Delta z}+{\cal O}(\Delta z)\right]
\label{e4.18}
\nonumber
\end{eqnarray}
and the relation obtained by shifting $m$ to $m+1$.

The leading asymptotic contribution to the eigenvalue condition in (\ref{e3.19})
in this case is balanced when again
\begin{eqnarray}
\kappa_{\rm c}={4\over 3-{D-2\over m+D-1}}\Bigm\vert_{D=4}={4(m+3)\over 3m+7}.
\label{e4.19}
\nonumber
\end{eqnarray}
The remaining terms to higher order are
\begin{eqnarray}
\Delta\kappa\sim -{24(m+1)(m+2)(m+3)\over (3m+7)^2} \Delta z \ln{\Delta z}
+{\cal O}(\Delta z).
\label{e4.20}
\nonumber
\end{eqnarray}
Inverting this relation then yields
\begin{eqnarray}
\Delta z\sim -{(3m+7)^2\over 24(m+1)(m+2)(m+3)}
{\Delta\kappa\over \ln{\Delta\kappa}} \left[1+{\cal O}\left({\ln\ln{1\over
\Delta\kappa}\over\ln{\Delta\kappa}}\right)\right].
\label{e4.21}
\nonumber
\end{eqnarray}
{}Finally, we obtain a logarithmic scaling relation for the adsorption fraction:
\begin{eqnarray}
P(\kappa)\sim -{3m+7\over 3(m+1)(m+2)} {1\over \ln (\kappa-\kappa_{\rm c})}
\quad (\kappa\to\kappa_{\rm c}^+).
\label{e4.22}
\end{eqnarray}

The behavior of the adsorption fraction is summarized in Fig.~\ref{f2}, where we
have plotted the adsorption fraction $P(\kappa)$ for $0\leq D\leq 6$ and for
$1.25\leq\kappa\leq 2$ by solving numerically the eigenvalue equation in
(\ref{e3.19}). In this plot we chose $m=0$ because the critical phenomena
derived in this section are most prominent for small values of the radius $m$.

\subsection{Crossover Transition}
\label{ss4.5}

In the limit of large radius $m$ we intuitively expect that the attractive
boundary will be effectively planar on the length scale of monomer units. In
this limit the asymptotic behavior of the adsorption fraction near the critical
point should thus be linear. Hence, for $m\gg 1$ there must be a {\sl crossover}
region such that the scaling coefficient obtained for the adsorption fraction
$P(\kappa)$ changes from being dimensionally dependent and sensitive to the
curvature of the boundary to the value $1$, which is obtained for the case
$D=1$. To be precise, if we allow the binding potential to vary in a small
neighborhood above $\kappa_{\rm c}$,
$$\kappa=\kappa_{\rm c}+\Delta\kappa \qquad (\Delta\kappa\ll 1),$$
for some fixed radius $m\gg 1$, we find that the relations given in 
Eqs.~(\ref{e4.12}-\ref{e4.22}) hold when $\Delta\kappa\ll 1/m$. However, for 
$1\gg\Delta\kappa\gg 1/m$ linear scaling is obtained. Consequently, the 
crossover between these two regimes occurs for any $D>0$ when 
$\Delta\kappa={\cal O}(1/m)$ (aside from possible logarithmic corrections for 
even integer $D$), as stated in Eq.~(\ref{e1.2}).

To locate the crossover region analytically it is useful to study the eigenvalue
condition in (\ref{e3.19}) asymptotically in the limits $\Delta z\ll 1$, $\Delta
\kappa \ll 1$, and $1/m \ll 1$. There are three distinct cases to consider: (i) 
$\Delta z\gg m^{-2}$, (ii) $\Delta z\sim m^{-2}$, and (iii) $\Delta z\ll
m^{-2}$. As was found in Ref.~\cite{Bo} for the particular case $D=2$, we find
here that for all $D>0$ cases (i) and (ii) lead immediately to linear scaling.
For case (iii) we find to leading order that
\begin{eqnarray}
\Delta\kappa\sim {8\Gamma\left({D\over 2}\right)\over
3^{1+{D\over 2}}\Gamma\left(1-{D\over 2}\right)m} \left(m^2\Delta
z\right)^{1-{D\over 2}}-{16\over 3D}m\Delta z\qquad (0<D<2)
\label{e4.23a}
\end{eqnarray}
and
\begin{eqnarray}
\Delta\kappa&\sim& {8\Gamma\left(2-{D\over 2}\right)\over
3^{3-{D\over 2}}\Gamma\left({D\over 2}-1\right)m} \left(m^2\Delta
z\right)^{{D\over 2}-1}-{16\over 3(4-D)}m\Delta z\nonumber\\
&&\qquad\qquad -{64\over (D-4)^2(D-6)} m^3\Delta z^2 \qquad (D>2,D\not= 4,6,
\ldots).
\label{e4.23b}
\end{eqnarray}
For all $D>0$ we therefore find that the $D$-dependent scaling relations derived
in Subsections \ref{ss4.1} - \ref{ss4.4} hold for case (iii). The crossover
transition happens when all the terms in Eqs.~(\ref{e4.23a}) and (\ref{e4.23b})
are of equal order, that is, where case (iii) borders on case (ii). Hence, the
crossover occurs when $m^2\Delta z={\cal O}(1)$, which verifies
Eq.~(\ref{e1.2}).

\section*{ACKNOWLEDGEMENTS}
\label{s5}
Two of us, CMB and PNM, wish to thank the U.S. Department of Energy for
financial support under grant number DE-FG02-91-ER40628. SB also thanks the U.S.
Department of Energy for support under grant number DE-AC02-76-CH00016.

\figure{
Random walk on a lattice consisting of concentric cylindrical surfaces of unit
radii. Such a walk serves as model for a polymer growing at an attractive
cylindrical boundary such as a cell membrane with radius $m$ (thickened lines).
The polymer is initially grafted to the boundary and is growing to the right.
Every time a monomer gets added at the boundary, the polymer gains in potential
energy by an amount $\kappa$. The walk consists of $N=26$ monomer links, but
only $L=14$ random steps were required because every random step in the radial
direction is followed by a deterministic step in the axial direction. This
requirement ensures that the random walk advances exactly one unit in the axial 
direction for each random step.
\label{f1}
}

\figure{
Plot of the adsorption fraction $P(\kappa)$. For increasing $D<2$ the scaling 
exponent increases and the transition becomes weaker until for $D=2$ exponential
scaling is obtained. For increasing $D>2$ the scaling exponent decreases and the
transition itself becomes stronger again which is compensated for by an increase
in the critical binding potential $\kappa_{\rm c}$ that is required to bring
about the transition. At $D=4$ we observe a tricritical point with logarithmic
scaling, and for $D>4$ the transition is first order, indicated by a
discontinuity (green shaded region) in $P(\kappa)$ across the critical point.
\label{f2}
}


\begin{references}
\bibitem{BeBoM}
C.~M.~Bender, S.~Boettcher, and L.~R.~Mead, J.~Math.~Phys.~{\bf 35}, 368 (1994),
and C.~M.~Bender, S.~Boettcher, and M.~Moshe, J.~Math.~Phys.~{\bf 35}, 4941
(1994).

\bibitem{BeCoMe}
C.~M.~Bender, F.~Cooper, and P.~N.~Meisinger, submitted to Phys.~Rev.~E.

\bibitem{BeBoMe}
C.~M.~Bender, S.~Boettcher, and P.~N.~Meisinger, submitted to Phys.~Rev.~E.

\bibitem{AIP} 
See for example {\it Random Walks and Their Application in the Physical and
Biological Sciences}, eds.~M.~F.~Shlesinger and B.~J.~West (American
Institute of Physics, New York, 1984).

\bibitem{DesCJ} 
J.~Des Cloizeaux and G.~Jannink, {\it Polymers in Solution} (Clarendon, Oxford,
1990).

\bibitem{DeG} 
P.~G.~DeGennes, {\it Scaling Concepts in Polymer Physics} (Cornell, New York,
1979).

\bibitem{Freed} 
K.~F.~Freed, {\it Renormalization Group Theory of Macromolecules}, (Wiley,
New York, 1987).

\bibitem{Binder}
K.~Binder and K.~Kremer, in {\it Scaling Phenomena in Disordered Systems}, eds.
R.~Rynn and A.~Skjeltorp (Plenum, New York, 1985), K.~Binder, Adv.~Polymer Sci. 
{\bf 112} 181 (1994), and K.~Binder, Nucl.~Phys.~B (Proc.~Suppl.) 42 (1995).

\bibitem{Bax} 
R.~J.~Baxter, {\it Exactly Solved Models in Statistical Mechanics},
(Academic, London, 1982).

\bibitem{FloMo} 
P.~J.~Flory, J.~Chem.~Phys. {\bf 17}, 303 (1949); E.~W.~Montroll,
J.~Chem.~Phys.~{\bf 18}, 734 (1950).

\bibitem{Sokal}
R.~Fernandez, J.~Fr\"ohlich, and A.~D.~Sokal, {\it Random Walks, Critical
Phenomena, and Triviality in Quantum Field Theory}, (Springer, Berlin, 1992).

\bibitem{Priv}
V.~Privman, G.~Forgacs, and H.~L.~Frisch, Phys.~Rev.~B {\bf 37}, 9897 (1988),
and V.~Privman and N.~M.~\v Svraki\'c, {\it Directed Models of Polymers,
Interfaces, and Clusters: Scaling and Finite-Size Properties}, (Springer,
Berlin, 1989).

\bibitem{BoMo}
S. Boettcher and M. Moshe, Phys.~Rev.~Lett.~{\bf 74}, 2410 (1995).

\bibitem{Bo}
S. Boettcher, Phys.~Rev.~E {\bf 51}, 3862 (1995).

\bibitem{Tr}
For the special case $m=0$, the eigenvalue problem given here resembles closely 
the one that was treated in Ref.~\cite{BeBoMe}. In fact, setting 
$g_0=(2a'/z')g_1'$, $g_n=g_{n+1}'~(n\geq 1)$, $P_{\rm in,out}(n)=P_{\rm in,out}'
(n+1)~(n\geq 0)$, $\lambda=\lambda'+1$, $z=z'/2$, and 
$\kappa=a'(\lambda'+1)/(a'+z'\lambda'/2)$ transforms Eqs.~(\ref{e3.1}) into (the
primed version of) Eqs.~(4.1) of Ref.~\cite{BeBoMe}.

\bibitem{A+S} M.~Abramowitz and I.~A.~Stegun, {\sl Handbook of Mathematical
Functions} (National Bureau of Standards, Washington, 1964), chap.~22.

\end{references}
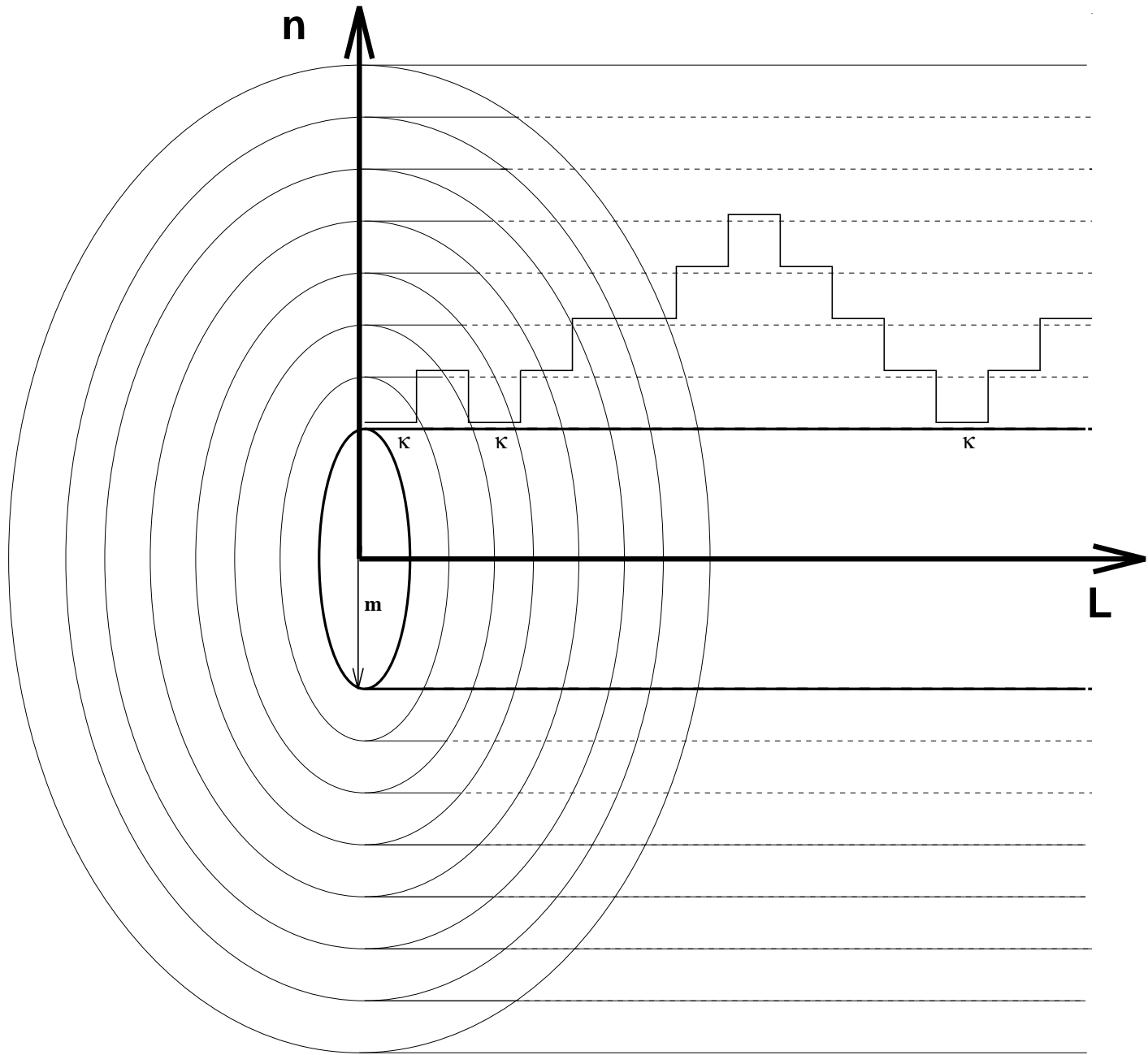
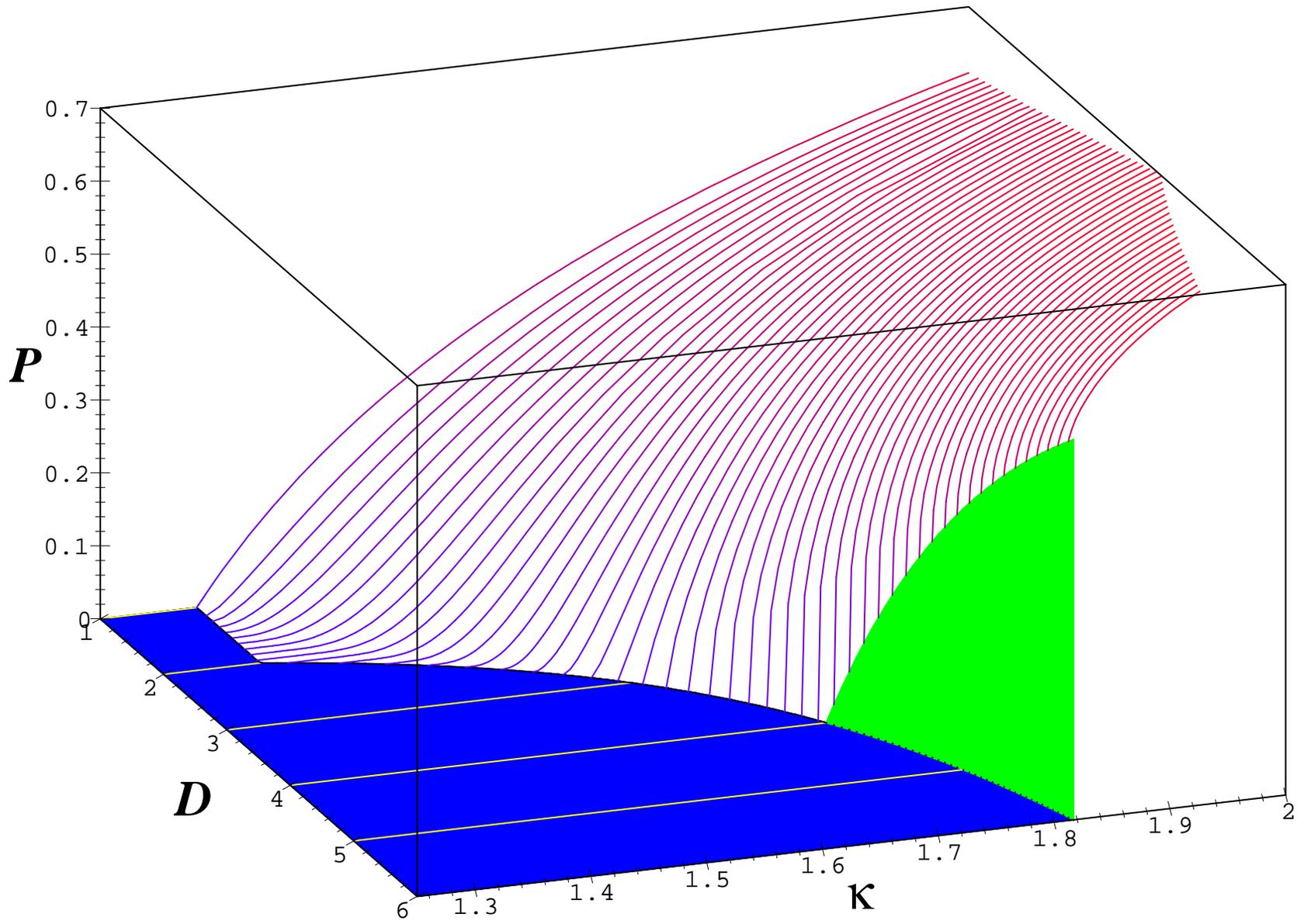
\end{document}